\documentclass[a4paper,onecolumn,12pt,draftclsnofoot]{IEEEtran}

\usepackage[font=footnotesize,skip=2pt]{caption}

\usepackage{framed}

\usepackage{cite}
\usepackage{graphicx}
\usepackage[cmex10]{amsmath}
\usepackage{amssymb}
\usepackage{booktabs}
\usepackage{threeparttable}
\usepackage{algorithm}
\usepackage{algpseudocode}
\usepackage{amsfonts}
\usepackage{xcolor}
\usepackage{color}

\usepackage{tabularx} %

\usepackage{dblfloatfix} %

\usepackage{soul}

\usepackage{cancel} %

\usepackage{verbatim}%

\usepackage{makecell}

\usepackage{empheq} %

\usepackage{underoverlap} %

\usepackage{hyperref}
\hypersetup{pdftex,colorlinks=true,allcolors=black}

\usepackage{enumitem} %
\usepackage{calc}

\usepackage{color}
\usepackage{array}
\usepackage{booktabs}
\usepackage{makecell}
\usepackage{rotating}
\usepackage{multirow}
\usepackage{colortbl}

\usepackage{setspace}

\usepackage{dashbox} %

\usepackage{textcomp}
\def\BibTeX{{\rm B\kern-.05em{\sc i\kern-.025em b}\kern-.08em
		T\kern-.1667em\lower.7ex\hbox{E}\kern-.125emX}}

\newenvironment{bluenote}{\color{blue}}

\newtheorem{remark}{\bf  Remark}

\newcommand{\HF}{hopping frequency}
\newcommand{\HFs}{hopping frequencies}

\newcommand{\mj}{\mathsf{j}}
\newcommand{\mC}{\mathsf{C}}

\newcommand{\myRound}[1]{\lfloor#1\rceil}
\newcommand{\myCeil}[1]{\lceil#1\rceil}

\newcommand{\myMod}[2]{\langle#1\rangle_{#2}}

\newcommand{\fft}[1]{{\mathrm{FFT}}_{#1}}
\newcommand{\ifft}[1]{{\mathrm{IFFT}}_{#1}}

\begin{document}

\title{\centering \setstretch{1}{\centering \parbox{160mm}{\centering A Low-Complexity Method for FFT-based OFDM Sensing}\\}}

\author{
	Kai Wu,
	J. Andrew Zhang,
	Xiaojing Huang and
	Y. Jay Guo

}

{}

\maketitle

\vspace{-2cm}

\begin{abstract}
	OFDM sensing is gaining increasing popularity in wideband radar applications as well as in joint communication and radar/radio sensing (JCAS). 
	As JCAS will potentially be integrated into future mobile networks where OFDM is crucial, OFDM sensing is envisioned to be ubiquitously deployed.
	A fast Fourier transform (FFT) based OFDM sensing (FOS) method was proposed a decade ago and has been regarded as a \textit{de facto} standard given its simplicity. In this article, 
	we introduce an easy trick --- a pre-processing on target echo --- to further reduce the computational complexity of FOS without degrading key sensing performance. 
	Underlying the trick is a newly disclosed feature of the target echo in OFDM sensing which, to the best of our knowledge, has not been effectively exploited yet.

\end{abstract}

\vspace{-0.5cm}
\begin{IEEEkeywords}
	\vspace{-0.3cm}
OFDM, radar sensing, multi-carrier, DFT, FFT, Decimation
\end{IEEEkeywords}

\IEEEpeerreviewmaketitle

\vspace{-0.5cm}

\section{Background and Motivation}\label{sec: background and motivation}

Orthogonal frequency-division multiplexing (OFDM) sensing has been a hot topic over the past decade.
Given its wideband nature, OFDM sensing is attractive in many radar applications that require high range resolution, such as radar imaging \cite{OFDM_imaging2015GRS}. 
Due to its high flexibility in waveform design and reconfiguration, OFDM 
is also a popular choice for software-defined radar \cite{OFDM_autonomousDriv2019microwaveMag}. Moreover, OFDM sensing is able to exploit the frequency diversity to survive interference-limited scenarios, e.g., automotive radar networks. In a similar way, OFDM is known to facilitate multiple-input and multiple-output (MIMO) radars in achieving high angular resolution \cite{BinYang_ofdmSPmagazine}. 
In fact, the initial motivation of introducing OFDM to radar sensing is to perform joint communication and radar sensing (JCAS) \cite{OFDM_1st2radar_levanon2000multifrequency}. 
The proliferation of wireless applications and ever-critical spectrum crowdedness make JCAS highly popular nowadays. 
In a recently proposed perceptive mobile network (PMN) \cite{Andrew_PMN2020TVTmag}, JCAS is integrated into ubiquitous mobile networks.
Thus, PMN is envisioned to achieve
ubiquitous sensing, or more specifically ubiquitous OFDM sensing, given the crucial role of OFDM in modern mobile networks \cite{book_ahmadi20195gNewRadio}.

A fast Fourier transform ({F}FT) {b}ased {O}FDM {s}ensing (FOS) method was proposed a decade ago by Sturm etc. \cite{DFRC_dsss2011procIeee}. Due to its low complexity and flexibility in accommodating classical radar sensing algorithms/theories, FOS has been regarded as a \textit{de facto} standard for OFDM sensing since its publication. 
In short, 
FOS collects $ M $ consecutive OFDM symbols, each having $ N $ sub-carriers.
After some pre-processing, two batches of FFTs are performed, one batch along sub-carriers and the other over symbols. 
A range-Doppler matrix (RDM) is then achieved, enabling target detection and parameter estimation. More details about FOS will be presented in Section \ref{sec: FOS}. 
Assume that $ M $ and $ N $ are powers of two. The computational complexity of the FFTs required for generating a single RDM, which also dominates the complexity of FOS, is given by $ \mathcal{O}\left( MN\log_2 N + N M\log_2 M \right) $, where
$ \mathcal{O}(x\log_2 x) $ is the complexity of the $ x $-point radix-two FFT $ (x=M,N) $ \cite{book_oppenheim1999discrete}, and $ M $ and $ N $ can take hundreds to thousands.

Despite that efficient FFTs can be used for generating RDM in FOS, we introduce in this article an easy trick, specifically an efficient pre-processing on the target echo, to further reduce the computational complexity by a substantial amount. Underlying the pre-processing is a newly disclosed feature --- the target echo in OFDM sensing contains high redundancy that can be removed and does not affect the sensing performance of major concern, e.g., maximum unambiguous range and velocity as well as their respective resolutions. 

The impact of our design can be profound. \textit{First}, it makes OFDM sensing promising to be implemented on a variety of Internet of things devices with relatively low computational power. This helps achieve ubiquitous sensing in the aforementioned PMN. \textit{Second}, since the time and resources saved from lowering the complexity of OFDM sensing can be used for making more prompt and comprehensive decisions, our new design can help reduce accidents and financial losses in time- and safety-critical applications.

\section{Signal Model of OFDM Sensing} \label{sec: signal model}

Consider a general JCAS scenario where OFDM communication symbols are also used for sensing through a full-duplex synchronized receiver (Rx) co-located with the transmitter (Tx). Provided that Tx and Rx are sufficiently isolated, no interference is from Tx to Rx directly. In addition,  
single-antenna Tx and Rx are employed to introduce the core idea that is independent of spatial information in theory. 

For the $ m $-th $ (m=0,1,\cdots,M-1) $ OFDM symbol, there are $ N $ data symbols to be transmitted, as denoted by $ s_m(n)~(n=0,1,\cdots,N-1) $. In OFDM, these $ N $ data symbols are multiplied onto $ N $ orthogonal sub-carriers which essentially are single-tone signals at center frequencies of $ n/T $. Here, $ T $ is the duration of the sub-carriers in the time domain. This further indicates that the bandwidth of the considered OFDM system is $ B=N/T $. 
Let $ T_{\mathrm{s}} $ denote the sampling time which takes $ T_{\mathrm{s}}=1/B=T/N $ in OFDM. Accordingly, the $ m $-th OFDM symbol can be expressed as a discrete function of time index $ k $, i.e., 
\begin{align} \label{eq: xmk = smn idft}
	x_{m}(k) = \frac{1}{N}\sum_{n=0}^{N-1} s_m(n)e^{\mj 2\pi n kT_{\mathrm{s}}/T} = \frac{1}{N}\sum_{n=0}^{N-1} s_m(n)e^{\mj 2\pi n k/N},~k=0,1,\cdots,N-1 .
\end{align}
From (\ref{eq: xmk = smn idft}), we see that multiplying data symbols with $ N $ orthogonal sub-carriers is equivalent to taking the $ N $-dimensional inverse DFT (IDFT) of the data symbols. 
In turn, taking the DFT of $ x_m(k) $ with respect to (w.r.t.) $ k $ can recover $ s_m(n) $. 

\begin{figure}[!t]
	\vspace{-0.8cm}
	\centering
		\begin{minipage}{70mm}
			\captionof{figure}{Illustrating the changes of signal timing in OFDM sensing, where CP is short for cyclic prefix and $ Q $ is the number of samples in a CP. The top signal, $ x_m(k) $ given in (\ref{eq: xmk = smn idft}), is the essential part of OFDM symbols. The middle signal, $ \tilde{x}_m(\tilde{k}) $ given in (\ref{eq: xmk CP added}), illustrates the CP-OFDM symbols to be emitted. The bottom signal, $ \tilde{y}_m(\tilde{k}) $ given in (\ref{eq: echo model}), is the baseband echo at the sensing Rx, where the delay of $ k_r $ samples account for the round-trip traveling from Tx to Rx.}	
			\label{fig: signal timing}	
		\end{minipage}\hfill
		\begin{minipage}{80mm}
			\centering
			\includegraphics[width=80mm]{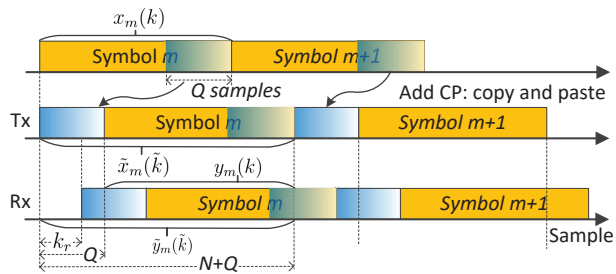}		
		\end{minipage}
	\vspace{-1cm}
\end{figure}

According to the circular shift property \cite{book_oppenheim1999discrete}, the DFT of any circularly shifted $ x_m(k) $ is still $ s_m(n) $ with extra phase shifts, 
which, based on (\ref{eq: xmk = smn idft}), can be translated into
\begin{align} \label{eq: circular shift property}
	x_{m}(\myMod{k-l}{N}) = \frac{1}{N}\sum_{n=0}^{N-1} \Big(s_m(n)e^{-\mj 2\pi ln/N}\Big)e^{\mj 2\pi n k/N} ,~k=0,1,\cdots,N-1,~\forall l
\end{align}
where $ \myMod{\cdot}{N} $ denotes modulo-$ N $. 
Since the sample delay $ l $ resembles the echo delay in the sensing Rx, it is implied by (\ref{eq: circular shift property}) that \textit{the sequence of $ s_m(n) $ can always be recovered from the target echo as long as a complete (circularly shifted) OFDM symbol is sampled}. 
To ensure this, a cyclic prefix (CP) is generally added to $ x_m(k) $ by copying the last $ Q $ samples and pasting them to the beginning of $ x_m(k) $; refer to Fig. \ref{fig: signal timing}.
Denoting the number of samples in the CP by $ Q $, the $ m $-th OFDM symbol becomes
\begin{align} \label{eq: xmk CP added}
	\tilde{x}_m(\tilde{k}) = x_m(\myMod{\tilde{k}-Q}{N}),~\tilde{k}=0,1,\cdots,N+Q-1,
\end{align}
which is obtained by plugging $ k=\myMod{\tilde{k}-Q}{N} $ into (\ref{eq: xmk = smn idft}). The timing relation between $ \tilde{x}_m(\tilde{k}) $ and $ x_m(k) $ is described in Fig. \ref{fig: signal timing}. 

Next, we build the signal model for target echos.
For illustration convenience and clarity, we model a single sensing target whose range, velocity and reflection coefficient are $ r $, $ v $ and $ \alpha $, respectively. We also assume that $ r $, $ v $ and $ \alpha $ keep constant over $ M $ OFDM symbols, as complied with the Swerling-I target fluctuation model \cite[Ch.7]{book_richards2010ModernRadarprinciples}. The round trip (from Tx to target then back to Rx) causes a delay of $ k_r=\myRound{2r/(\mC T_{\mathrm{s}})}  $ samples in the target echo, as compared with the transmitted OFDM symbol, where $ \myRound{x} $ rounds $ x $ to the nearest integer and $ \mC $ is the microwave propagation speed.
The target velocity incurs a Doppler frequency which can be calculated as $ \mu=2vf_{\mathrm{c}}/\mC $, where $ f_{\mathrm{c}} $ denotes the carrier frequency of the sensing system. Taking into account $ k_r $ and $ \mu $, the target echo can be modeled as
\begin{align}\label{eq: echo model}
	\tilde{y}_m(\tilde{k}) = \alpha g(\tilde{k}) \tilde{x}_m(\tilde{k}-k_r)e^{\mj 2\pi m \tilde{T}\mu }, ~\tilde{k}=0,1,\cdots,N+Q-1
\end{align}
where $ g(\tilde{k})=0 $ for $ \tilde{k}=0,1,\cdots,k_r-1 $ and $ g(\tilde{k})=1 $ for the remaining values of $ \tilde{k} $; and $ \tilde{T}=T+QT_{\mathrm{s}} $ denotes the time duration of a CP-OFDM symbol. The echo timing with reference to the emitted signal is exemplified in Fig. \ref{fig: signal timing}. Though noises are inevitable in any practical Rx, they are suppressed in (\ref{eq: echo model}) for brevity. Moreover, the ``hop-and-stop'' model \cite{book_richards2010ModernRadarprinciples} has been used for the Doppler effect by omitting the intra-symbol Doppler-related change.

\section{FFT-based OFDM Sensing (FOS)}\label{sec: FOS}

From Fig. \ref{fig: signal timing}, we see that the non-trivial part of $ \tilde{y}_m(\tilde{k}) $ contains a circularly shifted OFDM if $ k_r\le Q $ is satisfied.
Under the condition, remove the first $ Q $ samples of $ \tilde{y}_m(\tilde{k}) $, yielding $ \bar{y}_m(k) = \alpha x_m(\myMod{{k}-k_r}{N}) e^{\mj 2\pi m \tilde{T}\mu } $ for $ k=0,1,\cdots,N-1 $. 
By taking $ l=k_r $ in (\ref{eq: circular shift property}), the DFT of $ x_m(\myMod{{k}-k_r}{N}) $ w.r.t. $ k $ is $ s_m(n)e^{-\mj 2\pi nk_r/N} $. Since $ \alpha e^{\mj 2\pi m \tilde{T}\mu } $ is a coefficient independent of $ k $, the DFT of $ \bar{y}_m(k) $ w.r.t. $ k $ can be directly given by 
$ \breve{y}_m(n) = \alpha s_m(n)e^{-\mj 2\pi nk_r/N} e^{\mj 2\pi m \tilde{T}\mu } $. 
Removing the data symbol $ s_m(n) $ in $ \breve{y}_m(n) $ leads to \textit{the pre-processed target echo}, as given by
\begin{align} \label{eq: ymn data removed}
	y_m(n) = \alpha e^{-\mj 2\pi nk_r/N} e^{\mj 2\pi m \tilde{T}\mu }.
\end{align}
It is not difficult to recognize that $ y_m(n)~(\forall m) $ is a discrete single-tone signal w.r.t. $ n $, and $ y_m(n)~(\forall n) $ is such a signal w.r.t. $ m $.
The center frequencies of the two single-tone signals are related to the range and velocity of the target, and can be estimated from the following two-dimensional DFT of $ y_m(n) $, i.e., the aforementioned RDM,
\begin{align} \label{eq: Ybk 2D DFT for range and velocity}
 	Y_{b}(k) = \alpha \sum_{n=0}^{N-1} w_{N}(n) e^{-\frac{\mj 2\pi n k_r}{N} } e^{-\mj \frac{2\pi k n}{N}} \times \sum_{m=0}^{M-1} w_{M}(m) e^{\mj 2\pi m \tilde{T}\mu } e^{-\mj \frac{2\pi  b m}{M}} ,
\end{align}
where $ w_{N}(n) $ and $ w_{M}(m) $ denote window functions of lengths $ N $ and $ M $, respectively. 
From (\ref{eq: Ybk 2D DFT for range and velocity}), we obtain that the peak of $ |Y_b(k)| $ locates at $ k^{\dagger}=N-k_r $ and $ \tilde{b}^{\dagger}=M\tilde{T}\mu $. Since $ k_r=\myRound{2r/(\mC T_{\mathrm{s}})}  $ and $ \mu=2v/\lambda $, $ r $ and $ v $ can be estimated as
\begin{align}\label{eq: r hat v hat}
	\hat{r} \approxeq  (N-k^{\dagger})\mC T_{\mathrm{s}}/2,~\hat{v} \approxeq b^{\dagger}\mC \Big/(2Mf_{\mathrm{c}}\tilde{T}),
\end{align} 
where $ b^{\dagger} $ is a modified version of $ \tilde{b}^{\dagger} $ to account for negative velocity. In particular, we have $ b^{\dagger}=\tilde{b}^{\dagger} $ if $ \tilde{b}^{\dagger}\le M/2 $; otherwise, $ b^{\dagger}=\tilde{b}^{\dagger}-M $.

\vspace{-0.3cm}
\section{Our Design to Reduce Complexity of FOS} \label{sec: dfos}
We proceed to introduce an efficient design that further simplifies the computational complexity of FOS. To start with, we disclose a key feature of $ y_m(n) $.
Rewriting $ y_m(n) $ leads to
\begin{align}\label{eq: ymn rewritten}
	y_m(n)=\alpha e^{\mj 2\pi m \tilde{T}\mu } e^{-\mj 2\pi (nT_{\mathrm{s}})\frac{k_r}{NT_{\mathrm{s}}}} =\alpha e^{\mj 2\pi m \tilde{T}\mu }  e^{-\mj 2\pi (nT_{\mathrm{s}})\frac{k_r B}{N}}, 
\end{align}
where the last result is due to $ B=1/T_{\mathrm{s}} $. From the above expression, we see that the frequency of $ y_m(n) $ is $ k_r B/N $. 
As underlined Section \ref{sec: FOS}, FOS requires $ k_r\le Q $. This indicates that the bandwidth of $ y_m(n) $ is no greater than $ QB/N=B/D $, where $ D=N/Q $. In OFDM communication systems, $ Q\ll N $ is satisfied \cite{book_ahmadi20195gNewRadio}. Thus, we obtain the following:

\textbf{Echo Feature:} \textit{Provided that the maximum sample delay in target echo is no greater than the CP length and the CP length is much less than the sub-carrier number, the pre-processed target echo has a much smaller bandwidth of than an OFDM symbol.}

Employing the signal models established previously, the above feature can be interpreted as: provided $ k_r\le Q\ll N $, $ y_m(n) $ given in (\ref{eq: ymn data removed}) has a much smaller bandwidth than $ x_m(k) $ given in (\ref{eq: xmk = smn idft}). 
With the feature identified, we see that only $ 1/D $ of the whole frequency band contains useful information for sensing and the rest is filled with noises. Namely, $ y_m(k) $ given in (\ref{eq: ymn data removed}) can have considerably redundant information. To this end, we introduce:

 \textbf{An Easy Trick:} \textit{Decimate the pre-processed target echo, i.e., $ y_m(n) $ given in (\ref{eq: ymn data removed}), to remove redundancy and hence reduce signal samples along the $ n $-dimension, prior to sensing}.

\begin{remark}
	The above echo feature may not be difficult to observe. However, to the best of our knowledge, exploiting the feature to lower the complexity of OFDM sensing has been overlooked in the past decade. With the efficient decimation to be introduced shortly, the complexity of FOS can be reduced by orders of magnitude. In addition, as will be analyzed in Section \ref{sec: comparison between FOS and DFOS}, the decimation does not affect the key sensing performance, e.g., range/velocity resolution.
\end{remark}

\vspace{-0.3cm}

\subsection{Efficient Decimation} \label{subsec: decimation}

\begin{figure}[!t]
	\centering
	\vspace{-0.5cm}	
	\includegraphics[width=145mm]{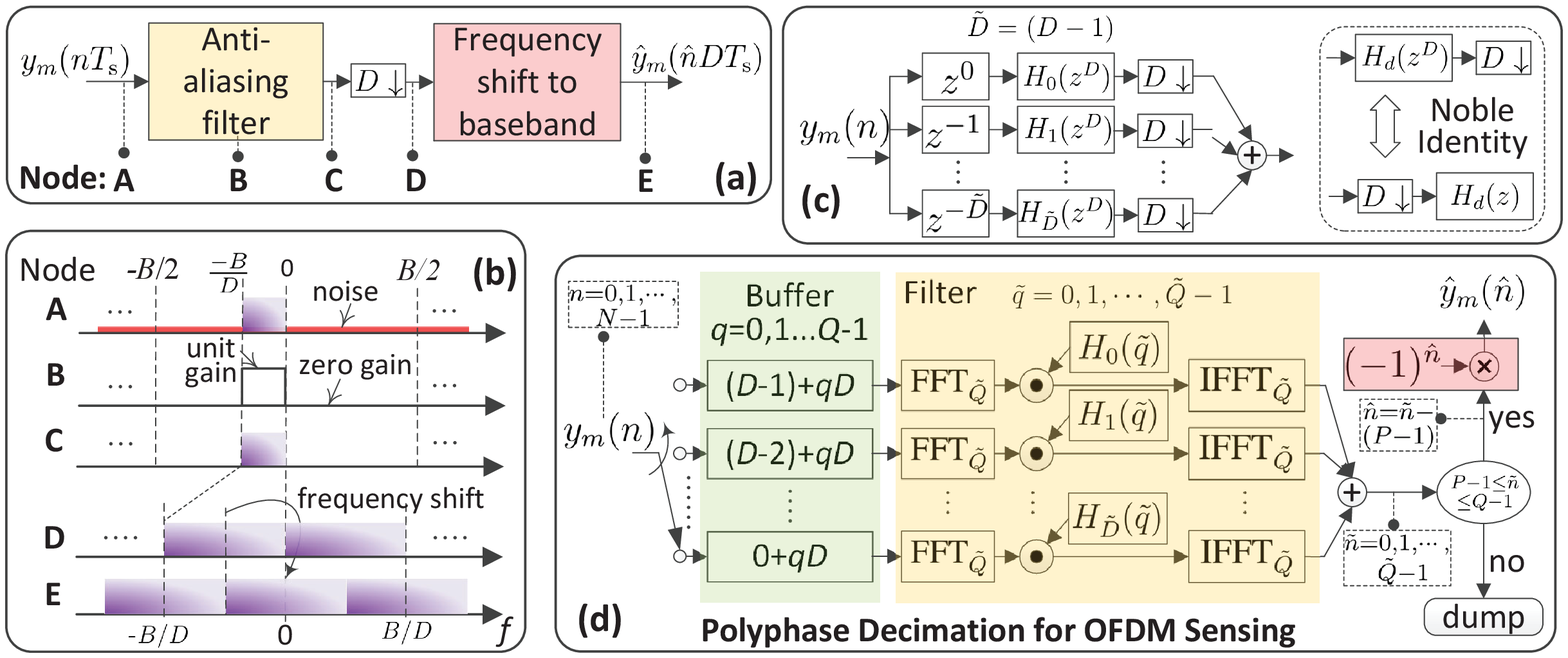}
	\caption{(a) Illustration of general steps for decimation; (b) spectrum features at different stages of decimation; (c) decomposing the anti-aliasing filter in Fig. \ref{fig: decimation}(a); (d) the polyphase structure-based decimation specifically tailored for OFDM sensing.}
	\label{fig: decimation}
	\vspace{-0.8cm}
\end{figure}

As seen from (\ref{eq: ymn rewritten}), $ y_m(n) $ is a bandpass signal with frequency band $ [-B/D,0] $. 
To decimate $ y_m(n) $ by the factor of $ D $, the following steps can be performed, as illustrated in Fig. \ref{fig: decimation}(a).
\begin{enumerate}[leftmargin=*]
	\item Anti-aliasing filtering: is performed on $ y_m(n) $ to suppress out-of-band interference and noises. 
	The passband of the filter is the same as that of $ y_m(n) $, while the stopband is given by $ [-B/2,B/2]\oslash[-B/D,0] $, where $ \oslash $ denotes set difference.
	The frequency spectrum of an ideal bandpass filter is shown in Node B of Fig. \ref{fig: decimation}(b). The signal spectrum before and after filtering is shown in Nodes A and C of Fig. \ref{fig: decimation}(b), respectively. 
	As ideally illustrated in Node C, out-of-band noises are totally removed, which is impractical but can be well approximated by designing the anti-aliasing filter with a large stopband attenuation.

	\item Downsampling: is denoted by ``$ D\downarrow $'' in Fig. \ref{fig: decimation}(a). It
	keeps every $ D $-th sample (starting from sample $ 0 $) and deserts others. After downsampling, the sampling frequency is reduced to $ B/D $, and the spectrum center becomes $ -B/(2D) $; see Node D of Fig. \ref{fig: decimation}(b).
	
	\item Frequency shifting: shifts the spectrum center of the downsampled signal to zero, which leads to the spectrum shown in Node E of Fig. \ref{fig: decimation}(b). 
\end{enumerate}
Above are the general steps of a bandpass decimation. 
By invoking the polyphase structure, the decimation can be implemented more efficiently.

At the core of the polyphase structure is the decomposition of the anti-aliasing filter. Consider an $ (L-1) $-order finite impulse response anti-aliasing filter. Let $ h(l) $ denote the $ l $-th $ (l=0,1,\cdots,L-1) $ filter coefficient.  
The $ z $-transform of $ h(l) $ can be expressed as \cite[Ch.6]{book_harris2004multirate}
\begin{align}\label{eq: Hz decompose}
	H(z) = \sum_{l=0}^{L-1}h(l)z^{-l} = \sum_{d=0}^{D-1} z^{-d} \sum_{p=0}^{P-1}h(d+pD) z^{-pD} = \sum_{d=0}^{D-1} z^{-d} H_d(z^D) ,
\end{align}
where the second equality is obtained by decomposing $ l=d+pD $ and the $ p $-related summation is denoted by $ H_d(z^D) $ in the last result. 
Note that $ L=PD $ is assumed in the above decomposition. The condition can be readily satisfied by specifying the filter order as $ (PD-1) $ when designing the anti-aliasing filter. Based on (\ref{eq: Hz decompose}), we see that the filter can be implemented in $ D $ parallel branches, as illustrated in Fig. \ref{fig: decimation}(c). The input signal $ y_m(n) $ goes into different branches simultaneously, and the outputs of the branch-filters, denoted by $ H_d(z^D)~(\forall d) $, are supposedly to be summed and then downsampled. But in Fig. \ref{fig: decimation}(c), we move the downsampler to before the summation and equivalently put a downsampler in each branch. Doing so allows us to invoke the notable identity, as illustrated in Fig. \ref{fig: decimation}(c), to exchange the orders of filter and downsampler in each branch. The 
order exchanging makes the delay block, $ z^{-d} $, adjacent to a downsampler. To this end, the samples to be filtered by the $ d $-th $ (\forall d) $ branch-filter become $ y_m(\tilde{D}-d+qD)~(q=0,1,\cdots,Q-1) $, where ``$ -d $'' reflects the $ d $-delay block in branch $ d $, $ \tilde{D}=(D-1) $ is added to sample indexes to ensure that the indexes are no less than zero, and $ qD $ is a result of the downsampler. 
Based on (\ref{eq: Hz decompose}), the coefficients of the $ d $-th branch-filter are $ h_{d+pD} ~(p=0,1,\cdots,P-1) $. 

The filter decomposition and the order exchanging illustrated above lead to the polyphase structure of bandpass decimation, as shown in Fig. \ref{fig: decimation}(d). 
In the figure, we use a buffer to collect continuous $ Q $ samples, i.e., $ y_m(\tilde{D}-d+qD)~(q=0,1,\cdots,Q-1) $ for the $ d $-th branch, and each branch-filter is implemented in the frequency domain due to the following relation 
\begin{align}
	h_{d+pD}\circledast y_m(\tilde{D}-d+qD) \equiv
	 \mathrm{IFFT}_{\tilde{Q}}\left\{ \fft{\tilde{Q}}\{h_{d+pD} \}\odot \fft{\tilde{Q}}\big\{y_m(\tilde{D}-d+qD) \big\} \right\}, \nonumber
\end{align}
where ``$ \circledast $'' denotes linear convolution, ``$ \equiv $''
means that the calculations on its two sides are equivalent, $ \ifft{\tilde{Q}} $ and $ \fft{\tilde{Q}} $ denote size-$ \tilde{Q} $ IFFT and FFT, respectively, and ``$ \odot $'' calculates the point-wise product. Note that the above equivalence requires $ \tilde{Q}\ge (P+Q-1) $.
For radix-2 (I)FFT, we can take $ \tilde{Q} $ such that $ \log_2\tilde{Q}=\myCeil{\log_2(P+Q-1)} $. 
Since each branch-filter produces $ (P-1) $ transient outputs and takes $ Q $ samples as input, the indexes of valid filter outputs are $ P-1,P,\cdots,Q-1 $. Thus, we keep the valid outputs and dump others, as shown in Fig. \ref{fig: decimation}(d).

Referring back to Fig. \ref{fig: decimation}(a), we are now at the last step of decimation, i.e, shifting the filtered and downsampled signal to the baseband. To differentiate with $ y_m(n) $, we use $ \hat{n} $ to denote the index of valid samples after downsampling, as also highlighted in Fig. \ref{fig: decimation}(d). 
Based on (\ref{eq: ymn rewritten}), the signal, after filtering and with transients removed, can be expressed as
\begin{align}
	\alpha e^{\mj 2\pi m \tilde{T}\mu } e^{-\mj 2\pi \hat{n}Dk_r/N}  = \alpha e^{\mj 2\pi m \tilde{T}\mu } e^{-\mj 2\pi \hat{n} k_r/Q} ,~\hat{n}=0,1,\cdots,Q-P. \nonumber
\end{align}
As a discrete function of $ \hat{n} $, the 
spectrum center of the above signal is now at $ \pi $, since the mean value of $ k_r $ is $ Q/2 $. 
According to the frequency shift property of Fourier transform, we know that an angular frequency shift of $ \pi $ can be equivalently realized by multiplying the time-domain sequence with $ e^{\mj \pi \hat{n}} = (-1)^{\hat{n}} $, which leads to the frequency shift block shown in Fig. \ref{fig: decimation}(d). Accordingly, the final output of the polyphase structure-based decimation is
\begin{align} \label{eq: ymn hat decimated}
	\hat{y}_m(\hat{n}) = \alpha e^{\mj 2\pi m \tilde{T}\mu } e^{-\mj 2\pi \hat{n} k_r/Q}\times e^{\mj \pi \hat{n}} = \alpha e^{\mj 2\pi m \tilde{T}\mu } e^{-\mj 2\pi \hat{n}  \frac{k_r+Q/2}{Q} }.
\end{align}

\vspace{-1cm}

\subsection{Decimation-based FOS (DFOS)} \label{subsec: dfos parameter estimation}

Similar to FOS reviewed in Section \ref{sec: FOS}, sensing can also be performed based on $ \hat{y}_m(\hat{n}) $, leading to the decimation-based FOS (DFOS). 
Taking the two-dimensional DFT of $ \hat{y}_m(\hat{n}) $ w.r.t. $ m $ and $ \hat{n} $ generates the below RDM (referred to as DFOS-RDM), which has a smaller size than the RDM given in (\ref{eq: Ybk 2D DFT for range and velocity}) (similarly referred to as FOS-RDM),
\begin{align} \label{eq: Ybk hat decimated RDM}
	\hat{Y}_{b}(\hat{k}) = \alpha \sum_{\hat{n}=0}^{Q-1} w_{Q}(\hat{n}) e^{-\frac{\mj 2\pi \hat{n} (k_r+Q/2)}{Q} } e^{-\mj \frac{2\pi \hat{k} \hat{n}}{Q}} \times \sum_{m=0}^{M-1} w_{M}(m) e^{\mj 2\pi m \tilde{T}\mu } e^{-\mj \frac{2\pi  b m}{M}} .
\end{align}
Identifying the peaks of $ |\hat{Y}_{b}(\hat{k})| $ along $ \hat{k} $- and $ b $-dimensions can estimate range and velocity, respectively. 
Assume that the 
$ \hat{n} $-related summation achieves the maximum
at $ \hat{k} = \hat{k}^{\dagger} $. It is easy to see from (\ref{eq: Ybk hat decimated RDM}) that the maximum is only achieved when $ k_r+Q/2+\hat{k}^{\dagger}=aQ $, where $ a $ takes an integer or zero.
Solving the equation subject to $ k_r\in [0,Q-1] $ yields,
\begin{align}
	\hat{k}_r = Q/2-\hat{k}^{\dagger},~\mathrm{if}~\hat{k}^{\dagger}\in [0,Q/2];~\hat{k}_r = 3Q/2-\hat{k}^{\dagger},~\mathrm{if}~\hat{k}^{\dagger}\in [Q/2+1,Q-1],
\end{align}
where $ \hat{k}_r $ denotes the estimate of $ k_r $. Comparing (\ref{eq: Ybk 2D DFT for range and velocity}) and (\ref{eq: Ybk hat decimated RDM}), we see that FOS and DFOS have the same velocity measurement.
To sum up, DFOS has the following range and velocity estimates, where $ \hat{v}  $ is given in (\ref{eq: r hat v hat}),
\begin{align} \label{eq: r hat v hat decimation}
	\hat{r}_{\mathrm{d}} = \hat{k}_rT_{\mathrm{s}}\mC/2,~\hat{v} \approxeq b^{\dagger}\mC \Big/(2Mf_{\mathrm{c}}\tilde{T}).
\end{align}

\vspace{-0.5cm}

\section{Comparison Between FOS and DFOS} \label{sec: comparison between FOS and DFOS}

In this section, we compare FOS and DFOS from numerous aspects, through which the advantages and disadvantages of introducing the efficient decimation are analyzed.

\textbf{Computational Complexity:} \textit{DFOS has a much smaller computation complexity 
	than FOS, provided $ Q\ll N $.
} (Note again that $ Q\ll N $ is readily satisfied in mobile communication systems \cite{book_ahmadi20195gNewRadio}.)
From Section \ref{sec: FOS}, the computational complexity of FOS is dominated by that of computing the two-dimensional RDM. 
This part of complexity has been given in Section \ref{sec: background and motivation}, i.e., $ \mathcal{O}\left( MN\log_2 N + N M\log_2 M \right) $, which equals to $ \mathcal{O}\left( MN\log_2 (MN) \right) $ by basic logarithmic laws. 

DFOS has two parts of computations: the two-dimensional FFT for generating DFOS-RDM and decimation. 
Like FOS, the first part of computation has the complexity of $ \mathcal{O}\big( MQ\log_2(MQ) \big) $. 
According to Fig. \ref{fig: decimation}(d), the computational complexity of the polyphase decimation is dominated by the first column of FFTs and the third columns of IFFTs. Their complexity is given by $ \mathcal{O}\left( 2D\tilde{Q}\log_2\tilde{Q} \right) $, since the first (third) column has $ D $ numbers of $ \tilde{Q} $-size FFTs (IFFTs). 
By designing the anti-aliasing filter such that $ P\ll Q $, we can take $ \tilde{Q}=2Q $ in the polyphase decimation, where $ Q $ is often an integer power of two in practice. To this end, $ \mathcal{O}\left( 2D\tilde{Q}\log_2\tilde{Q} \right) $ becomes $\mathcal{O}\left( 4DQ\log_2(2Q) \right) $. Note that $ 4DQ\log_2(2Q) $ is much smaller than $ MQ\log_2(MQ) $, since $ M $ can take several hundreds while $ D $ is around ten. Thus, the computational complexity of DFOS is approximately $ \mathcal{O}\big( MQ\log_2(MQ) \big) $.

\textbf{Processing Gain:} \textit{FOS and DFOS have approximately the same processing gain which is defined as the difference between the SNR in the RDM, i.e., in (\ref{eq: Ybk 2D DFT for range and velocity}) and (\ref{eq: Ybk hat decimated RDM}),
	and the SNR in the pre-processed target echo, i.e., in (\ref{eq: ymn data removed}).
 }Let $ \gamma $ denote the SNR of $ y_m(n) $ given in (\ref{eq: ymn data removed}). Although noises are not explicitly shown in the signal models, the SNR change is easy to track. FOS-RDM is obtained from a two-dimensional DFT of $ y_m(n) $, and hence the SNR in FOS-RDM becomes $ MN\gamma $. Namely, the processing gain of FOS is given by $ MN $.

DFOS decimates $ y_m(n) $ first. The decimated version $ \hat{y}_m(\hat{n}) $ given in (\ref{eq: ymn hat decimated}) has the SNR of $ D\gamma $, since the decimation with factor $ D $ does not change signal power while reduces the noise power by $ D $ times. The two-dimensional DFT performed in (\ref{eq: Ybk hat decimated RDM}) improves the SNR to $ M(Q-P+1) D\gamma\approx MN\gamma $, where $ M(Q-P+1)\approx MQ $ 
and the approximation is valid given $ P \ll Q $. We see that the processing gain of DFOS is approximately $ MN $.

\begin{remark}
	The impact of $ P $ on DFOS can be non-trivial. 
	For instance, as $ P $ increases, a higher quality filter can be obtained (e..g, one with lower passband ripple, stronger stopband attenuation and narrower transition bandwidth); however, a lower processing gain, as given by $ M(Q-P+1) $, is yielded. 
	Analytically, it is difficult to tell which of the following dominates: the SNR improvement earned by a better filter or the SNR degrading caused by the reduced processing gain. To this end, we resort to simulation next. 
\end{remark}

Figs. \ref{fig: snrvs p gain}(a) and \ref{fig: snrvs p gain}(b) illustrates that, as $ P $ increases from $ 1 $ to $ 50 $, the SNR in DFOS-RDM first increases, then plateaus, and next decreases. The same pattern is seen for both small and large values of $ \gamma $. From this observation, we conclude that the SNR in DFOS-RDM can be maximized by properly setting $ P $. For the OFDM system configured in Figs. \ref{fig: snrvs p gain}, the maximum is achieved at $ P=16 $.
Using this value, we compare in Fig. \ref{fig: snrvs p gain}(c) the SNR in DFOS-RDM with that in FOS-RDM, as $ \gamma $ increases. We see that the SNRs achieved by FOS and DFOS are almost identical in the whole region of $ \gamma $. Note that the difference between the $ y $-axis and $ x $-axis is the processing gain. Thus, the results in Fig. \ref{fig: snrvs p gain}(c) validate that FOS and DFOS have approximately the same processing gain.

\begin{figure}[t!]
	\vspace{-0.7cm}
	\centering
	\includegraphics[width=140mm]{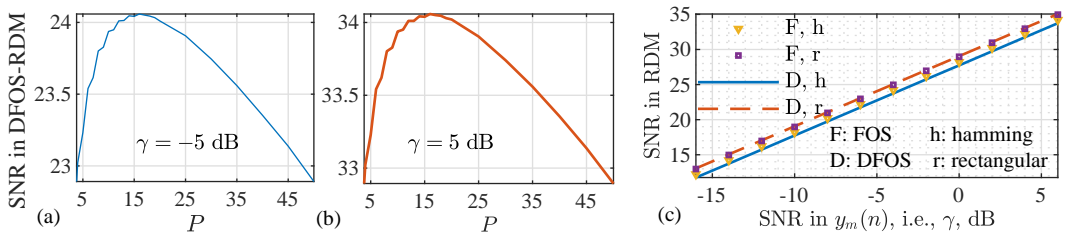}
	\caption{Illustration of SNR in DFOS-RDM versus $ P $ in (a) and (b); and (c) a comparative illustration of the SNR in RDM of both FOS and DFOS versus $ \gamma $, the SNR in (\ref{eq: ymn data removed}). The OFDM sensing system is parameterized according to \cite[Tab.2]{DFRC_dsss2011procIeee}, where the carrier frequency is $ f_{\mathrm{c}}=24 $ GHz, the number of sub-carriers is $ N=1024 $, the symbol duration is $ T=11\mu $s, the CP length is $ Q=128 $ (thus the decimation factor is $ D=N/Q=8 $), and the quadrature phase shift keyig (QPSK) modulation is adopted for communications.
A single target is set in this simulation, where the $ r\in\mathcal{U}_{[0,200] } $ (m), $ v\in\mathcal{U}_{[-110,110] } $ (m/s) and $ \alpha=1 $. Here $ \mathcal{U}_{[a,b]} $ denotes a uniform distribution between $ a $ and $ b $. The results shown in the figures are averaged over $ 2\times 10^4 $ independent trials. For simplicity, we set the number OFDM symbols in each trial as $ M=1 $. 
}
	\label{fig: snrvs p gain}
	\vspace{-1cm}
\end{figure}

\textbf{Range and Velocity Measurement:} \textit{FOS and DFOS share the same maximum unambiguous range/velocity; they also have the same range/velocity resolution.}
In terms of velocity, the above statement is because the decimation does not incur any change to Doppler-related information, as manifested in (\ref{eq: Ybk 2D DFT for range and velocity}) and (\ref{eq: Ybk hat decimated RDM}). Based on (\ref{eq: Ybk 2D DFT for range and velocity}),  
the range of Doppler frequency that can be unambiguously estimated is $ \mu\in [-\frac{1}{2\tilde{T}},\frac{1}{2\tilde{T}}] $, where $ \frac{1}{\tilde{T}} $ resembles the sampling frequency along the Doppler dimension. Since the number of samples is $ M $, the Doppler frequency resolution is $ \Delta_{\mu}=\frac{1}{\tilde{T}M} $. Given the relation $ \mu=2v/\lambda $, we obtain the range of unambiguous velocity, i.e., $ v\in[-\frac{\lambda}{4\tilde{T}},\frac{\lambda}{4\tilde{T}}] $, and the velocity resolution, as given by $ \Delta_{v}=\frac{\lambda}{2\tilde{T}M} $.

It terms of ranging, we see from (\ref{eq: Ybk 2D DFT for range and velocity}) and (\ref{eq: Ybk hat decimated RDM}) that the range estimation is turned into the problem of identifying $ k_r $ in both FOS and DFOS. 
Since $ k_r(=\myRound{2r B/\mC}) $ is independent of the sampling rate (or range dimension) in different RDMs, its estimate remains the same for FOS and DFOS in theory.
As illustrated in Section \ref{sec: FOS}, $ k_r\le Q $ is required for OFDM sensing. 
Let $ R $ denote the maximum unambiguous detectable range. 
Solving $ 2R B/\mC=Q $, we obtain $ R=\frac{\mC Q}{2B} $, for both FOS and DFOS. We see from (\ref{eq: Ybk 2D DFT for range and velocity}) and (\ref{eq: Ybk hat decimated RDM}) that the resolution of $ k_r $ detection is unit one for both methods, and hence the range resolution, denoted by $ \Delta_{r} $, can be solved from $ 2\Delta_{r} B/\mC=1 $, leading to $ \Delta_{r}=\frac{\mC}{2B} $.

\textbf{Windowing Effect:} \textit{For ranging, FOS can achieve a better windowing effect than DFOS in the sense that FOS has a narrower range mainlobe than DFOS given the same attenuation of peak sidelobe, while for velocity measurement, the two methods have the same windowing effect.} The reason is because the decimation in DFOS reduces the number of samples, hence the window length, along the range dimension (compared with those of FOS), while the decimation does not affect the velocity dimension.

The range and velocity measurements using FOS and DFOS are compared in Fig. \ref{fig: rdms range and vel cuts}. 
From Figs. \ref{fig: rdms range and vel cuts}(a) and \ref{fig: rdms range and vel cuts}(b), we see high similarity between the RDMs of the two methods. This validates the efficacy of the newly introduced decimation. It is noteworthy that DFOS reduces the complexity of generating the RDM shown in the figure by almost an order of magnitude, compared with FOS. This can be readily validated by substituting the parameter settings in the above complexity analysis. 
Fig. \ref{fig: rdms range and vel cuts}(c) compares the range cuts between FOS and DFOS. We see that DFOS has a slightly wider mainlobe than FOS, which is caused by different window lengths. 
Fig. \ref{fig: rdms range and vel cuts}(d) compares the velocity cuts of the two methods. As expected, our design does not affect the velocity measurement.

\begin{figure}[t!]
	\vspace{-0.8cm}
	\centering
	\includegraphics[width=140mm]{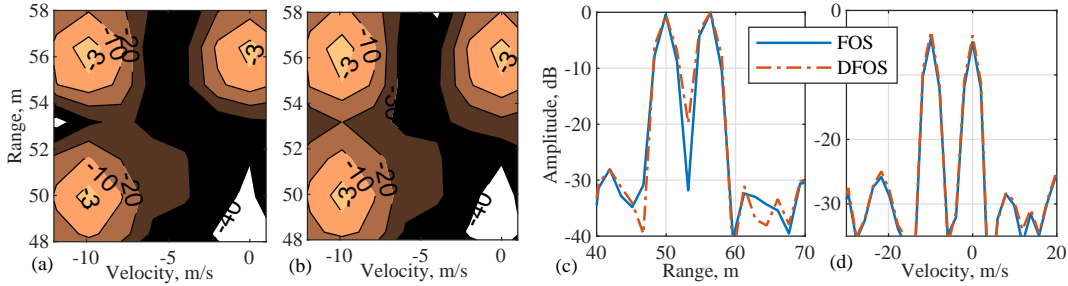}
	\caption{Illustration of target detection, where the contour plot of FOS-RDM is give in (a), that of DFOS-RDM in (b), the range cuts at $ v=-10 $m/s are shown in (c), and the velocity cuts at $ r=56 $m in (d). 
		The same OFDM system as set in Fig. \ref{fig: snrvs p gain} is used here, except that the number of OFDM symbols is $ M=256 $ and the hamming window is used in (\ref{eq: Ybk 2D DFT for range and velocity}) and (\ref{eq: Ybk hat decimated RDM}) for both range and velocity measurements. In addition, three targets are set here. Their ranges and velocities are $ [50,56,56] $m and $ [-10,-10,0] $m/s, respectively. All three targets have $ \alpha=1 $.}
	\label{fig: rdms range and vel cuts}
	\vspace{-1cm}
\end{figure}

\vspace{-0.5cm}

\section{Conclusion}
With 
a pre-processing introduced, DFOS is proposed for OFDM sensing in this article, substantially reducing the computational complexity of the famous FOS. An efficient polyphase structure-based implementation of the pre-processing is illustrated. Comparisons between FOS and DFOS are made from numerous aspects, which are validated by simulation results. While the key sensing performance remains the same between FOS and DFOS, a better windowing effect can be achieved by the former. 
A detailed analysis of the impact of the degraded windowing effect on sensing and how to solve it deserve more research effort, which is left for future work.

\vspace{-0.3cm}

\ifCLASSOPTIONcaptionsoff
\newpage
\fi

\bibliographystyle{IEEEtran}
\bibliography{IEEEabrv,./bib_JCAS.bib}

\end{document}